\documentclass[final,3p,times,twocolumn]{elsarticle}
\usepackage{amssymb,graphicx, epsfig}

\newcommand{\AmS}{{\protect\the\textfont2
  A\kern-.1667em\lower.5ex\hbox{M}\kern-.125emS}}

\journal{Astroparticle Physics}

\begin{document}

\begin{frontmatter}

\title{Astrophysical models for the origin of the positron ``excess''}

\author{Pasquale D.~Serpico}
\address{LAPTh, UMR 5108, 9 chemin de Bellevue - BP 110, 74941 Annecy-Le-Vieux, France}

\begin{abstract}
Over the last three years, several satellite and balloon observatories have suggested intriguing features in the cosmic ray lepton spectra. Most notably, the PAMELA satellite has suggested an ``anomalous'' rise with energy of the cosmic ray positron fraction. In this article, we summarize the global picture emerging from the data and recapitulate the main features of different types of explanations proposed.  The perspectives in testing different scenarios as well as inferring some astrophysical diagnostics from current/near future experiments are also discussed.
\end{abstract}

\begin{keyword}
Cosmic ray electrons and positrons; Pulsars; Supernova remnants.\flushright Preprint LAPTH-032/11
\end{keyword}
\end{frontmatter}

\section{Introduction}

Probes of very different nature, both cosmological and astrophysical, suggest that the Universe is CP-asymmetric, containing negligible amounts of antimatter. Concerning our Galactic environment, this is confirmed by the strong dominance of the matter with respect to the antimatter component in cosmic rays (CRs). The small amounts of positrons and antiprotons detected in CRs are attributed to byproducts of collisions of  CRs in the rarefied  interstellar medium (ISM), a process whose probability can be inferred from the grammage deduced by other secondary/primary ratios like the boron-to-carbon ratio, B/C.  In turn, this permits to use these species as additional handles on CR propagation and interactions in the ISM as well as to perform consistency checks of the models.

Turning the argument around, under the assumption of purely secondary production, the positron flux can be computed in a relatively robust way, of course subject to
some uncertainties coming from primary fluxes, cross sections and propagation parameters, see e.g.~\cite{Delahaye:2008ua}.
Deviations from these expectations have even been proposed as indirect signatures of dark matter (DM)  in the Galactic halo (see the pioneering article~\cite{Silk:1984zy} and refs. to it):  In most popular models where DM is made of (meta)stable weakly interacting massive particles (WIMPs), their annihilations or decays in the Galactic halo produce GeV-TeV energy Òcosmic raysÓ equally abundant in particles and antiparticles. Since antiprotons constitute only 10$^{-4}$ of the proton flux and positrons only  $\sim {\cal O}(10\%)$ of the electron flux, the lower background makes antimatter the preferred channel for searches of WIMP signatures in the flux of charged cosmic rays.

It is thus not surprising that a great excitation has followed the release of new data on the positron fraction $e^{+}/(e^{+}+e^{-})$ by the PAMELA collaboration~\cite{Adriani:2008zr,Adriani:2010ib}, as well as the important complementary information on $e^{+}+e^{-}$~\cite{Torii:2008xu,:2008zzr,Aharonian:2008aa,Abdo:2009zk,Aharonian:2009ah,Ackermann:2010ij} and $e^{-}$ spectra~\cite{Adriani:2011xv}. In particular, the positron fraction in the cosmic ray spectrum measured by PAMELA appears to begin climbing quite rapidly between $\sim$7 GeV till at least $\sim$100 GeV. Although previous experiments such as HEAT~\cite{Beatty:2004cy}   and AMS-01~\cite{Aguilar:2007} were already hinting to a possible deviation from basic expectations, the PAMELA data make this evidence striking thanks to a much higher statistics and extend it over a wider energy range. A recent, independent confirmation of this behaviour has been published by Fermi~\cite{:2011rq}.

The concern was raised (see e.g.~\cite{Schubnell:2009gk}) that this signal could be mimicked by misidentified protons in the instrument, if a rejection power no better than $\sim {\cal O}(10^{-4})$  was achieved by PAMELA. Since then, the collaboration has extensively argued (see~\cite{Adriani:2010ib}) that their rejection power---both based on test beam data and on in-flight data---can be estimated at the $10^{-5}$ level or lower, thus strongly disfavouring proton contamination as a plausible cause. In the following, we shall assume that the data reflect a real feature. This will be anyway independently confirmed by AMS-02~\cite{AMS-02}, which was launched on May 16, 2011 and is successfully operating onboard of the International Space Station. Note that a more subtle artifact signal may follow if the detected particles are truly positrons, but rather deriving from locally produced secondary particles, for example in the pressurized container of the PAMELA detector. Judging from the simulations reported in~\cite{bruno09}, it appears that the number of such events is negligible.  

\begin{figure}[!htb]
\begin{center}
\begin{tabular}{c}
\epsfig{figure=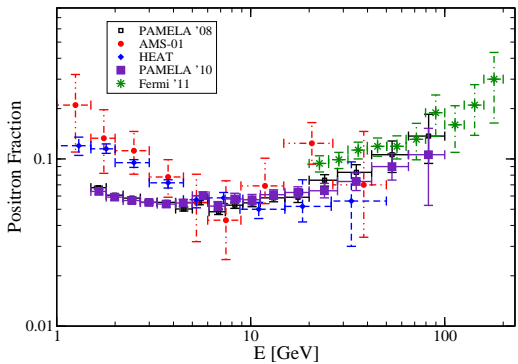,width=1.0\columnwidth}
\end{tabular}
\end{center}
\vspace{-0.9pc} \caption{The positron fraction vs. energy measured by PAMELA 2008~\cite{Adriani:2008zr} (statistical errors only), PAMELA 2010~\cite{Adriani:2010ib} (systematical and statistical errors summed in quadrature), and Fermi 2011~\cite{:2011rq}, compared
with previous results from HEAT (combined data)~\cite{Beatty:2004cy} and AMS-01~\cite{Aguilar:2007}. }\label{plot1}
\end{figure}

\begin{figure}[!htb]
\begin{center}
\begin{tabular}{c}
\epsfig{figure=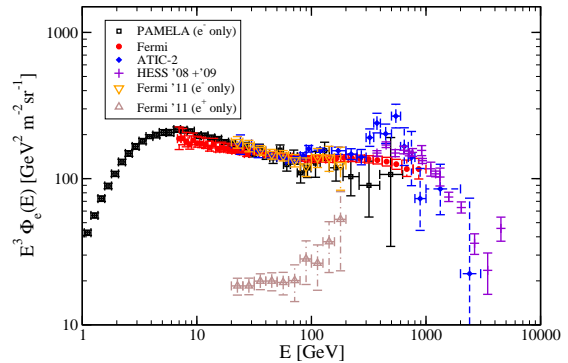,width=1.0\columnwidth}
\end{tabular}
\end{center}
\vspace{-0.9pc} \caption{Electron plus positron fluxes published since 2008 by PPB-BETS~\cite{Torii:2008xu} (statistical errors only), ATIC~\cite{:2008zzr} (statistical errors only),
HESS~\cite{Aharonian:2008aa,Aharonian:2009ah} (statistical errors only) Fermi~\cite{Abdo:2009zk,Ackermann:2010ij}, as well as the $e^{-}$ spectra recently
published by PAMELA~\cite{Adriani:2011xv} and the $e^{-}$ and $e^{+}$ spectra recently presented by Fermi~\cite{:2011rq}. For Fermi and PAMELA data, systematics and statistical errors are summed in quadrature.}\label{plot2}
\end{figure}

Here we have no ambition of reviewing in depth the observational situation of leptonic CR data (a recent review covering these aspects has been presented in~\cite{Fan:2010yq}.) In particular, for a detailed discussion of the error budget we refer to the original publications. Yet, since we shall refer to recent observations extensively, in Fig.~\ref{plot1} we report the PAMELA and Fermi positron fraction data together with a couple of previous determinations at low energy, while in Fig.~\ref{plot2} we report the electron+positron flux as published by several experiments since 2008 together with the electron-only flux recently presented by PAMELA and the electron-only and
positron only fluxes recently presented by Fermi. 

This review article is structured as follows: in Sec.~\ref{excess} we sketch why these data appear to require a source of primary positrons (or, most likely, of electrons and positrons). In Sec.~\ref{DM} we briefly summarize why DM is unlikely to be involved in the explanation of the phenomenon. In  Sec.~\ref{pulsars} we outline the basic arguments in favour of pulsars as sources, as well as some open problems in the astrophysics of these objects. In  Sec.~\ref{SNR} we describe the main alternative, namely (hadronic) production and acceleration in Supernova Remnants (SNRs), which are considered the standard sources of hadronic CRs~\cite{Blandford:1987pw}. In Sec.~\ref{alternatives} we briefly mention alternative astrophysical explanations and also explain the prominent role of local sources. Finally, in Sec.~\ref{Conclusions} we summarize the current status and outline some perspectives for the future.

\section{An excess... with respect to what?}\label{excess}
The claim of ``anomaly'' in the positron fraction has been initially presented by the PAMELA collaboration as well as in many interpretation papers as an excess
with respect to the no-reacceleration model prediction of~\cite{Moskalenko:1997gh}, although with some caveat concerning the
absence of error estimate. At face value, it is unclear to what extent one can consider the positron data as ``anomalous'', unless
some firm statement on the positron fraction {\it prediction} can be made. In~\cite{Serpico:2008te}, the anomaly was reformulated as a  {\it spectral feature}
(namely a rising curve) rather than as an excess problem. 

Let us illustrate this point more quantitatively. 
The positron fraction can be written as
\begin{equation}
f(E)\equiv\frac{1}{1+(\Phi_{e^{-}}/\Phi_{e^{+}})}\approx \frac{1}{1+\kappa\, E_{\rm GeV}^\varrho}\,,
\end{equation}
where the fluxes $\Phi_i$ refer to the ones at the top of the atmosphere, and a simple fit to the data above 10 GeV (to minimize
the charge-dependent solar modulation effects) yields $\varrho=-0.38\pm0.06$ for the original PAMELA analysis~\cite{Adriani:2008zr}, or  $\varrho=-0.23\pm0.04$ for the alternative analysis presented in~\cite{Adriani:2010ib}. At the same time, a fit of the PAMELA electron data over the corresponding energy interval provides $\Phi_{e^{-}}\propto E^{-3.23\pm0.02}$. As a consequence, one infers an absolute positron
spectrum which goes as $\Phi_{e^{+}}\propto E^{-2.85\pm0.06}$ (respectively $\Phi_{e^{+}}\propto E^{-3.00\pm0.04}$). Now, the proton flux measured by PAMELA itself in the similar energy-range is $\Phi_{p}\propto E^{-2.82}$~\cite{Adriani:2011cu}, and it is this hadronic CR spectrum which enters as source for the ISM positron one~\footnote{The recent indication by CREAM~\cite{Ahn:2010gv} and PAMELA~\cite{Adriani:2011cu} that Helium spectra are about $\sim 0.1$ harder than proton ones and that a hardening takes place in the 10$^2\div 10^3$ GeV range at several hundreds of GeV does not change qualitatively the conclusion.}. In a pure diffusive propagation framework, secondary/primary yields indicate that diffusion {\it alone} would steepen the spectrum by about $\delta\simeq 0.4$,  see e.g.~\cite{DiBernardo:2009ku} for more quantitative details. More importantly, TeV-leptons are subject to large energy losses, steepening their spectral index significantly more: for example by $(1+\delta)/2\approx 0.7$ in a two-zone diffusion model (where sources are embedded in a thin disk immersed in the thick halo~\cite{Acr90}) or  by $\approx 1$ in a leaky box model where radiative losses fully dominate the propagation.  For illustration, the recent numerical analysis in~\cite{Lavalle:2010sf} predicts a secondary positron spectrum roughly declining as $\Phi_{e^{+}}\propto E^{-3.4}\div E^{-3.5}$ in the range of interest, clearly yielding a $e^+$ fraction inconsistent with the data. In view of some concerns about theoretical assumptions entering these considerations raised for example in~\cite{Katz:2009yd}, it is worth noting that the only ``theoretical'' argument is to assume that, as a result of the diffusion-loss process in the propagation through the ISM, there is no viable (i.e. consistent with known physics and cosmic ray astrophysics) mechanism which can steepen the spectral index of 10-100 GeV positrons with respect to their production by $\sim 0.2$ or less, i.e. by less than the purely diffusive steepening effect. Needless to say, should such a kind of alternative solution be found, it would constitute a much greater breakthrough than the addition of primary sources~\footnote{Note that in the ``nested leaky box model'' the interpretation of the data is completely different,  but one assumes as starting point that in fact there are {\it two components} of ``secondary'' CRs like Li, Be, B, $e^+$ and $\bar{p}$ (see~\cite{Cowsik:2010zz} for details). These are produced both in the ISM and inside high density ``cocoons'' surrounding sources, successively escaping in the ISM. We do not address here advantages and problems of this model, but note that in that case the need for a source contribution to ``secondaries'' is built-in as {\it assumption}. Of course, debating on the need of a source component is only meaningful in diffusion or leaky box models trying to fit the data with only ISM secondaries.}. 
Of course, the need for an additional component can be confirmed in more refined statistical analyses, see~\cite{Trotta:2010mx,Liu:2011re,Auchettl:2011wi}, but we believe that it is important to stress the simplicity and generality of the basic argument.
  
Since we argued that the problem is inherent to the positron spectrum, present data strongly disfavour alternative scenarios where the anomalous positron fraction is actually attributed  to a non-conventional shape of the electron flux, e.g. due to the sum of a diffuse disk population and a local one as in~\cite{Shaviv:2009bu}. While it appears natural that a homogeneous approximation for lepton propagation breaks down at high-energies---even below the TeV scale---and likely future high-statistics data will be sensitive to inhomogeneity effects, quantitatively they appear too small to explain the anomaly in present observations (see also the recent analysis~\cite{Stockton:2011ks} suggesting that the  inhomogeneities assumed in~\cite{Shaviv:2009bu} are too extreme.) We come to the conclusion that the simplest and most conservative explanation of the observed phenomena is that a population of primary positron  (or more likely of positron and electron) accelerators exist.

This conclusion is gaining further support from recent Fermi measurements of the positron flux (i.e. separating it from the total $e^{-}+e^{+}$ flux) via deflections in the Earth magnetic field, reported in~\cite{:2011rq}, revealing a positron spectrum certainly harder than $E^{-3.0}$ in the range of interest. One component fits (i.e. only ISM production) of Fermi data fail to reproduce both the total $e^{-}+e^{+}$ flux as well as the $e^{+}$ flux~\cite{priv}, providing additional support for: a) dismissing proton contamination as main source of the effect; b) the need for an additional electron-positron source to provide a satisfactory fit of the data.

Once accepting that in order to explain the data one needs a primary source of positrons in addition to the secondary ones produced in the ISM, the pressing question 
becomes: what is (are) this (these) source(s)?  Many answers have been proposed in the literature and we shall try to review them in the following, with an emphasis on the likely astrophysical explanations.

\subsection{Another anomaly?}\label{Modulation}
Although way less discussed than its high-energy counterpart, by simple inspection of Fig.~\ref{plot1} one notices a discrepancy at low-energy between PAMELA data and
previous experiments like AMS-01 and HEAT. In a certain sense, this is even more puzzling: an experimental agreement is
found in an energy range which disagrees with the simplest expectations, while disagreement among experiments seems to appear where no new
source effect is expected!

More seriously, this trend might even raise the question if the entire measurement is plagued by some systematic problem. It is not the purpose of
the present review to discuss this issue in details. However, it suffices to say that it is generally accepted that no inconsistency is present
at low energy, the effect being due to {\it charge dependent} solar modulation. Some investigations of this effect have been presented e.g. in~\cite{GSicrc09},
and it has been extensively studied in~\cite{difelice} which we address for further details. Since PAMELA has been taking data for half a solar cycle,
it should be possible by now to see this trend to vary when time-binning is used, as well as an anti-correlation with a similar effect in antiproton/proton data.
It is clear that with more and more high-quality data available over long period of times, antimatter studies are becoming a more and more refined tool for heliospheric studies,
see for example the forecasts for AMS-02 presented in~\cite{Bobik:2010pe}.

\section{Is DM a good bet for explaining the CR lepton features?}\label{DM}

At first sight, an ``excess'' in the positron fraction is qualitatively expected in DM annihilation  models, hence explaining the
excitement generated by these data among particle physicists. Unfortunately, typical predictions from WIMP dark matter
annihilation fail to reproduce the data in three main aspects: 
\begin{itemize}
\item[i)] First of all, the peculiar signature for an exotic origin of the signal would rather be
a spectral edge (more or less sharp, depending on the final state) after an initial rise; this drop back to the expected background at high energies is not observed, neither in the $e^{+}$ fraction nor reflected as a dip in the $e^{+}+e^{-}$  data~\footnote{The so-called ``ATIC peak''~\cite{:2008zzr} has not been confirmed by later Fermi~\cite{Abdo:2009zk,Ackermann:2010ij} and HESS data~\cite{Aharonian:2008aa,Aharonian:2009ah},
although no consensus exists in the community for an explanation of the disagreement.}.  

\item[ii)] Insisting in attributing the rise to DM, the normalization of its
contribution seems to be surprisingly large, compared with typical expectation for a S-wave annihilating
thermal relic matching the observed dark matter abundance: i.e. in terms of annihilating cross sections, $\langle \sigma v\rangle \gg\langle \sigma v\rangle_{\rm th, S-wave}\approx$ 1 pb. 

\item[iii)] Such a large yield of $e^{\pm}$ pairs should imply  large yields of high-energy particles  in other channels; in particular one expects
anomalies in antiprotons, gamma-rays and possibly neutrinos, while no anomaly has been revealed yet in this respect.
\end{itemize}
The first kind of problem can be overcome by pushing the mass of the DM $m_X$ to values sufficiently higher than the range explored by PAMELA, perhaps to the few TeV scale
where the lepton fluxes appears to fall (see Fig. 2).   Since the differential spectrum from DM scales as $m_X^{-2}$, this exacerbates the second problem and, as we shall comment upon,  also affects the third one. In turn, the second problem can be ÒcuredÓ in three qualitatively different ways:
\begin{enumerate}
\item By attributing the high flux to a nearby clump of DM (as proposed for example in~\cite{Hooper:2008kv}), which appears however very unlikely in most structure formation scenarios
and given existing gamma-ray constraints~\cite{Brun:2009aj}.
\item By accepting that $\langle \sigma v\rangle \gg$ 1 pb and trying to derive that in a consistent model. The most popular scenario is to invoke a Sommerfeld
enhancement~~\cite{Sommerf31},  i.e. the effect of  non-perturbative enhancement of the annihilation due to long-range---compared to the TeV$^{-1}$ scale---attractive forces: this would be active in the low-velocity $v\simeq 10^{-3}\,c$ environment of our Galaxy, but would be inoperative in the early universe, thus making the standard calculation of relic abundance still approximately valid. In general, to accommodate such a mechanism one
needs to introduce additional light scales/particles with special coupling properties, at the expense of large fine-tunings. Alternatives include: i) non-thermal dark matter candidates---whose production does not happen by freeze-out, rather e.g. from decays of other species/condensates; hence $\langle \sigma v\rangle$ can be large---ii) some Breit-Wigner resonant enhancement; iii) to abandon standard cosmological evolution altogether. Of course, these options still require some ad hoc tuning of parameters, since the relic abundance of the particle is decoupled from its indirect signatures. 
\item By considering a mechanism of production of the positrons which does not involve annihilations at all, rather DM decay. Unfortunately, the lifetime $\simeq 10^{26}\,$s required to fit the data is not predicted in these scenarios, rather obtained a posteriori, although its scale might be justified in GUT models. The same ``decoupling'' from production affecting solutions of type 2 is of course plaguing these scenarios.
\end{enumerate}

It is impossible to review here the different models proposed: a brief (and by now incomplete) review on the subject was attempted in~\cite{[31]}. In any case, the absence
of signals in channels other than the $e^\pm$ and the type of spectrum observed pose serious constraints on
the particle physics details of the models. 
Since the presentation of first diffuse data by Fermi, more and more stringent limits have been derived from gamma rays, starting from the exploratory studies~\cite{Cirelli:2009dv,Papucci:2009gd} to the more recent ones in~\cite{Dugger:2010ys,Zaharijas:2010ca}, which add to stringent bounds following from cosmic microwave background data (for a recent study see~\cite{Galli:2011rz}), antiprotons~\cite{Donato:2008jk}, or radio data~\cite{Pato:2009fn,Crocker:2010gy}.  Models that evade these constraints are quite contrived, see e.g.~\cite{Cirelli:2010nh,Finkbeiner:2010sm}, and are unlikely to survive a closer scrutiny since they involve many approximations. A possibly incomplete list of these is: a) they {\it do not} provide {\it a fit} to diffuse gamma-ray data, i.e. they implicitly need astrophysical sources anyway, whose  role would be to tighten the constraints; b) for heavy DM, they neglect effects related to electroweak bremmsthralung~\cite{Kachelriess:2007aj,Bell:2008ey,Kachelriess:2009zy,Ciafaloni:2010ti,Bell:2010ei}, which have been shown to impact constraints (see for example~\cite{Bell:2011eu,Ciafaloni:2011sa,Bell:2011if,Garny:2011cj}); c) when engineering ``light mediators'' to have a mass just below some hadron mass to forbid significant  annihilation into gammas and antiprotons, three-body final states  are neglected, although in some concrete examples this has been proven to be a bad approximation~\cite{Yaguna:2010hn,Choi:2010jt}; d) in the case where a light scalar  $\phi$ with relatively large coupling $g_X$ to DM is the particle emitted in annihilation final states, a significant correction is expected to the two body channel. For example, the 3-$\phi$ emission with a soft $\phi$ should yield relatively large ``infrared'' log corrections to cross section, of the kind $g_X^2/(4\pi)\,\log(m_{X}/m_{\phi})$. This in turn softens the spectrum and enhances the ``boost'' required to fit a given dataset of leptonic data at high-energy, besides making the model more vulnerable to low-energy CR constraints.

Note that if one sticks to the conventional WIMP paradigm for DM, the present data by themselves are rather neutral to it, since typical expectations for DM signals in antimatter fall a couple of orders of magnitude below the fluxes observed. We do not treat the DM explanations for CR leptons any further, but we note that the burst of activity following the data had at least two healthy effects on the community: on one hand, many reflections have been developed on hidden assumptions in the standard WIMP paradigm, and the spectrum of theoretical possibilities have been broadened. On the other hand, a reality check has been imposed on the whole program of ``indirect DM detection''; this has revealed clearly that, facing the poorly known astrophysics,  barring some exceptions no indirect evidence for DM can be robustly found outside a well-defined and overconstrained framework such as the vanilla WIMP paradigm. Current data are already limited by systematic uncertainties on the astrophysics, rather than by statistics (see e.g. the case of the diffuse emission~\cite{Zaharijas:2010ca} or Galactic Center~\cite{Vitale:2009hr} one in gamma-rays.) It is fair to conclude that, until a better understanding of the astrophysical sources is achieved, most antimatter signatures of WIMPs are far from robust and a ``blind'' search via multi-messenger data mining appears poorly motivated. A new chapter might be eventually opened in DM phenomenology if the search for CR traces could be formulated as an {\it a priori} problem, namely with input from a discovery at colliders and/or direct detection experiments underground. In that case, in principle {\it correlated} predictions among many channels/directions/energies can be made which are unlikely to be mimicked by {\it uncorrelated} astrophysical sources.

\section{Pulsars (and Pulsar Wind Nebulae)}~\label{pulsars}
The idea that pulsars might be associated with the production of cosmic ray electrons/positrons is quite old, see~\cite{Arons:1981,Harding:1987,Boulares:1989}, and has been periodically reconsidered in the past decade, see e.g.~\cite{ZC01,Grimani07}.  The reason is that, being pulsars identified with fast rotating magnetized neutron stars, a large electric field is induced which can extract electrons from the star surface: in fact these effects are so strong that a pulsar rotating with angular velocity $\Omega$ is not living in vacuo, rather it is surrounded  up to a distance known as light radius $r_L=c/\Omega$ by a comoving plasma configuration called ``magnetosphere'' (see e.g.~\cite{reviewbook1,reviewbook2,reviewbook3}.)

In turn, the stripped electrons lose energy via curvature radiation (plus additional processes) while propagating far from the star along the magnetic field lines, and the emitted photons are so energetic that an electron-positron pair can be formed in the intense neutron star magnetic field. Numerous QED processes induce a multiplicative cascade which populates the magnetosphere of the pulsar with pairs.  The pairs produced in the magnetosphere, together with the Poynting flux emanating from the pulsar, form a relativistic magnetized wind evolving in a rich environment: since the pulsar is born from the collapse of a massive star (core collapse supernova) it lies initially well within the ejecta of its progenitor, in turn surrounded by the supernova blast wave propagating in the ISM. When the cold, magnetized relativistic wind launched by the star hits the non-relativistically expanding ejecta, a shock wave system forms in the impact: the outer one propagates in the ejecta, while a reverse shock propagates back towards the star. The latter is known as termination shock, where the wind is slowed down, its bulk energy dissipated and turned into that of a relativistically hot, magnetized fluid, which then shines as a Pulsar Wind Nebula (PWN).

If we denote with  $B_s$  the surface magnetic field and $R_s$  the radius of the neutron star, from Faraday's law one can naively estimate  a potential drop of up to $\varphi\simeq \omega B_s\,R_s^2 \sim10^{16}\,$V available for accelerating $e^{\pm}$ in the magnetosphere. Those pairs (which are more closely associated with the coherent pulsating radio and gamma emissions)  suffer at least adiabatic energy losses while reaching the termination shock. Nonetheless,  at the termination shock a relatively large fraction (a few tens of percent) of the wind bulk energy is converted into accelerated pairs, which then radiate a broad-band photon spectrum, extending from radio frequencies to multi-TeV gamma-rays, through synchrotron and Inverse Compton processes (see~\cite{Gaensler:2006ua} for an overview.)

It is unclear what the ultimate fate of these pairs is:  while propagating far from the termination shock advected together with the toroidal magnetic field lines, one expects them to be confined within the cavity created by the wind and to progressively lose energy. The fraction of them escaping in the ISM depends probably on astrophysical details. 

Given this broad picture, can one attribute the signal inferred from the charged lepton data to the PWN contribution? There are several elements which suggest that  a positive answer is possible if not likely, which we briefly address below.

\subsection{Energy budget}
A clear upper limit to the total energy into $e^{\pm}$ is given by the pulsar spin-down power integrated over time, i.e. the energy loss
corresponding to the slowing rate of rotation which in the standard theory provides virtually all of pulsar energy. In sufficiently general form, 
the decrease of the rotation frequency $\Omega$($=2\pi/P$, $P$ being the period) can be written
\begin{equation}
\dot{\Omega}=-\alpha\Omega^n\,,\label{dotom}
\end{equation}
yielding a  spin-down luminosity ($I$ being the momentum of inertia)
\begin{equation}
{\cal L}=|\dot E| =I \Omega |\dot\Omega|= \alpha\,I\,\Omega^{n+1}\,.
\end{equation}

The solution of Eq.~(\ref{dotom}) yields
\begin{equation}
\Omega(t)=\frac{\Omega_0}{\left(1+\frac{t}{\tau_0}\right)^{\frac{1}{n-1}}}\Rightarrow{\cal L}(t)=\alpha\,I\frac{\Omega_0^{n+1}}{\left(1+\frac{t}{\tau_0}\right)^{\frac{n+1}{n-1}}}\,,
\end{equation}
where we introduced the characteristic timescale
\begin{equation}
\tau_0\equiv[\alpha(n-1)\Omega_0^{n-1}]^{-1}\,.
\end{equation}

The special case of magnetic dipole corresponds to $n=3$ and $\alpha=5\,B_s^2 R_s^4/(8\,M_s\,c^3)$, 
where $M_s$ the mass of the neutron star. By integrating the
luminosity over time one gets 
\begin{eqnarray}
&&E_{tot}=\frac{1}{2}I\, \Omega_0^2\approx\\ 
&&\approx 2.2\times 10^{46} \left(
\frac{M_s}{1.4 M_\odot} \right)  \left(\frac{R_s}{10\,{\rm km}} \right)^2   \left(\frac{\Omega_0}{\rm Hz}\right)^2 \, \rm erg\,,\nonumber
\label{eq:energymax}
\end{eqnarray}
which amounts typically to $E_{tot}\simeq 10^{49}\,$erg or more mostly converted into a magnetized, relativistic wind in a time comparable with
$\tau_0$. Since one expects a rate $R_{CC}$ of about 2 core-collapse SN per century in our Galaxy (following for example from $^{26}$Al gamma-ray data~\cite{Diehl:2006cf}), the maximum luminosity injected is of the order of
\begin{equation}
{\cal L}_{\rm max}\approx 6.3\times 10^{39}\,{\rm erg\,s}^{-1}\frac{R_{CC}}{2\,{\rm century}^{-1}}\frac{E_{tot}}{10^{49}\,{\rm erg}}\,,
\end{equation}
to be compared to what needed to fit the data which is about one to two orders of magnitude lower, see e.g.~\cite{Hooper:2008kg,Kamae:2010ad}.

\subsection{Spectral shape}

{\it Observationally}, the spectrum of  radiation from several PWNe requires a spectrum of electrons and positrons which has a broken power-law shape: the break happens at Lorentz factor of $10^5\div 10^6$---hence at the ${\cal O}$(100) GeV scale---with a very hard power-law below the break, of the kind $\simeq E^{-1.5}$, while significantly softer (softer than $E^{-2}$) above the break (see e.g.~\cite{Gaensler:2006ua,Bucciantini:2010pd}).  It is important to emphasize that this is only based on data (mostly radio-to-X ray data) and does not rely on theoretical arguments. The hard spectral index matches what needed to fit the data~\cite{Hooper:2008kg,Kamae:2010ad,Malyshev:2009tw,Barger:2009yt}.

So, on general grounds, one can conclude that: i) pulsars have more than enough rotational energy to explain the normalization of the cosmic ray lepton features. ii)~Evidently, a significant part of this energy is converted into kinetic energy of lepton pairs at the termination shock of the PWNe, with the right spectral properties to match the high energy positron fraction as well as $e^{-}+e^{+}$ energy spectra. Also, it is worth adding that observations from Fermi-LAT instrument show clearly that pulsars (with or without the identification of PWNe) are the most abundant population of Galactic objects to shine in the GeV gamma-ray band~\cite{Collaboration:2010ru}. Including their contribution to any account of the leptonic CR sources is thus a requirement for any realistic description of Galactic CR sources. Reversing the argument, it is somewhat surprising that until recently---when sufficiently precise leptonic data in the $100-1000$ GeV have become available---one could fit the data without necessarily including them!

\subsection{Theoretical open problems}
Of course, the above arguments do not settle the many theoretical issues that still remain in the interpretation of these objects, for details see for example~\cite{Kirk:2007tn}. What PWNe observations seem to imply is a relativistic wind mostly made of electron-positron pairs (so mostly carrying particle kinetic energy), although it originates as a Poynting flux-dominated outflow  in the magnetosphere close to the pulsar. How does this conversion take place? In particular, how and where does the large multiplicity of pairs implied by observations originate?
The broken power-law spectrum is also very peculiar: even for the high-energy part which is steeper than $E^{-2}$, in apparent agreement with expectations from
 diffusive shock acceleration at a relativistic shock, the highly relativistic quasi-perpendicular nature of the shock should make acceleration inefficient, see e.g.~\cite{Sironi:2009jw}. The very hard low-energy spectrum is even more puzzling, since it implies a situation where most non-thermal particles lie at low-energy, but the energy is carried to a large extent by the few particles close to the high-energy break. To name but one of the associated puzzles: why the (expected) bulk of thermal low-energy particles acting as reservoir for the accelerated ones has never been identified in PWN?
 
Still, several interesting ideas have been put forward which address at least partially some of the issues. A cyclotron absorption acceleration mechanism~\cite{cyclabs} has been proven in simulations~\cite{Amato:2006ts} to be able to produce the desired spectra and efficiencies if a substantial fraction of the pulsar wind energy is carried by protons or other ions, namely if they are energetically dominant even if their number is negligible with respect to leptons~\footnote{This is of course conceptually possible, given the large rest mass ratio $m_p/m_e\simeq 1836$.}. Interestingly, it could also present preferential acceleration
of positrons thanks to helicity-matching of the waves onto which they scatter, which are generated by the (positively charged) protons.
But the required high Lorentz factor of the wind, although to some extent theoretically motivated, seems at odd with recent simulations as e.g. in~\cite{Bucciantini:2010pd}.
  
Another conceivable mechanism invokes magnetic reconnection of the striped magnetic field of the wind at low latitude around the pulsar rotational equator, which at the termination shock would convert into particle kinetic energy~\cite{Lyubarsky:2008yi}. The shape of the particle spectrum is however dependent on poorly understood details and it is difficult to confirm it observationally at present. Yet, recent two- and three-dimensional particle-in-cell simulations seem to support this kind of scenario~\cite{Sironi:2011zf}.

It is worth emphasizing once again that, although it is mandatory to solve these issues in order to put the theory of PWN acceleration and non-thermal emission on more solid ground, to some extent these problems are less directly linked to the contribution that PWNe can give to CR leptons seen at Earth. Since the spectrum of PWN is relatively well-known, more crucial to answering cosmic-ray questions are issues rarely addressed such as the escape probability of the accelerated pairs into the ISM, simply because they are way less relevant for ``photon astronomy''. In that respect, an interesting proposal has been done in Ref.~\cite{Blasi:2010de}, where it is argued that the so-called Pulsar Bow Shock Nebulae (i.e. high-velocity pulsars which have escaped their host SNR) might release their high-energy leptons directly in the ISM, a few tens of thousand years after their birth. In general, since the efficiency of conversion of spin-down luminosity into CR leptons is quite high, the effective efficiency of $\cal{O}$(1\%) would follow mostly  from the smaller residual energy fraction available after the escape from the SNR, rather than from a relatively inefficient conversion. This argument is also at ease with the non-dipolar spin-down laws---$2\lesssim n\lesssim 3$ in Eq.~(\ref{dotom})---found in the few pulsars for which the second derivative $\ddot\Omega$ has been measured ($n=-\Omega\,\ddot\Omega/\dot\Omega^2$).
  
\section{Production in SNRs}\label{SNR}
Usually, one neglects the production of ``secondaries'' within SNRs with the argument that the
probability of interaction for a primary species like protons is small over the lifetime of the accelerator, and in any case smaller than the probability
of interaction during the diffusion time in the ISM. For example, for a typical inelastic cross section of 30 mb one finds an interaction probability of about 10$^{-2.5}$ in a typical ISM density of 1 particle cm$^{-3}$ and for a representative lifetime of a source of 10$^5$ yrs. This has to be compared with a probability more than one order of magnitude larger to
undergo the same process during the propagation in the diffusive halo, which takes up to ${\cal O}$(10$^7$) yr, although in a medium whose density is in average much smaller.

However, this argument does not take into account that the spectrum resulting from the ISM production is steeper 
than the source one due to diffusion effects (roughly $E^{-2.7}$ vs $E^{-2.2}$), because it depends on the steady state spectrum of primaries rather than their injection one.  Hence, one expects that at sufficiently high energy the source production terms can make a significant contribution to the overall secondary spectrum, as already noted in~\cite{Berezhko:2000vy} for gamma-rays and more recently discussed for antiprotons~\cite{Blasi:2009bd}. Naively, an approximately flat secondary/primary ratio is expected as a result.

Recently, it was noted in~\cite{Blasi:2009hv} that an even more peculiar behaviour may follow when realizing that the secondary production takes place in the same region where cosmic rays are being accelerated. With the reasonable assumption that the diffusion coefficient $D$ of relativistic particles near the accelerating SNR shock grows with energy,  the region from which particles can return to the shock and undergo further acceleration gets larger and larger with energy. If, at the same time, the maximum energy attainable $E_{\rm max}$ is not limited simply by the naive diffusion timescale $t_{\rm diff}\sim D/v^2$ ($v$ being the shock velocity scale), one may obtain growing secondary/primary ratios, which can fit the current data. Note that a non-trivial behaviour of $E_{\rm max}$ is needed anyway to allow particles to reach knee energies via diffusive shock acceleration, since the naive estimate from typical ISM values for the diffusion coefficient would yield much smaller energies, as noted a long time ago~\cite{LC83}! A lot of attention has been paid lately on the proposal that this problem could be overcome via a non-resonant amplification of magnetic field by CRs themselves~\cite{Bell2004}, which seems to be confirmed (although not unambiguously) by the relatively thin X-ray rims seen in several SNRs, e.g.~\cite{Caprioli:2008st}. In general, a firm prediction is hampered by the very uncertain evolution of the magnetic field (amplification, damping, etc.) especially at the late stage of SNR evolution which are thought to contribute the most to ``low-energy'' (i.e. sub-TeV) CRs.
Additionally, non-trivial time and space evolution effects are expected (both for the plasma dynamics around the shock front as well as the background ``target'' matter), as for example recently tried to model phenomenologically in~\cite{Caprioli:2011tp}. Finally, the details of the injection and escape of CRs in/out of SNRs are yet unclear.
So, the steady state effective description attempted in~\cite{Blasi:2009hv} is clearly a simplification of the picture only meant as effective model. Not surprisingly, time-dependent simulations where most of these simplifications are retained but  $E_{\rm max}$ is fixed to its naive expectation find ``only'' a flat contribution to the Secondary/Primary ratio~\cite{Kachelriess:2011qv}, while they confirm the rising behaviour found in~\cite{Blasi:2009hv} if $E_{\rm max}$ is left free to assume larger values.  Fortunately, this scenario appears to be testable independently of the fact that a firm theory is not yet available. In fact, if  $E_{\rm max}$  (or an equivalent quantity) is treated as a free parameter to fit e.g. the positron data, a clear prediction  is obtained for associated observables like the antiproton flux or antiproton/proton ratio~\cite{Blasi:2009bd} or secondary/primary nuclei~\cite{Mertsch:2009ph}.
While the former predicts a rise at energies just beyond those currently explored, the nuclear ratios show a more ambigous picture, with some species like the titanium-to-iron ratio  from ATIC showing evidence for a rise, and others like the B/C data from other experiments favouring a decreasing function. Most likely this issue will be settled in the near future, especially thanks to AMS-02 data (see for example the forecasts in~\cite{Oliva:2008zz}).

Of course, the ``textbook'' model of SNR sources of CR via a strong, non-relativistic, planar shock is a ``zeroth order'' approximation of what happens in nature. Not all SNe are identical, and significant qualitative differences may exist between relatively light and relatively heavy progenitors. Massive stars like Red Supergiants ($15 M_\odot\lesssim M\lesssim 25 M_\odot$) and Blue Supergiant (or ``Wolf-Rayet'') stars ($M\gtrsim 25 M_\odot$)  explode into their stellar wind, which is magnetized, relatively dense and helium and metal-enriched from exposing the deeper layers of the star through mass ejections. The  magnetic field topology is expected to be radial in a polar cap, and tangential over most of the remaining solid angle. From the different conjectured forms of the diffusion tensor and acceleration mechanism in these two environments, it was argued that a harder ($E^{-2}$) component from the polar cap would eventually emerge over the steeper ($E^{-7/3}$) component which is numerically dominant at lower energies~\cite{Biermann:1993wy}.  
By adjusting the normalization parameters, it was shown in~\cite{Biermann:2009qi} that this allows one to fit leptonic data, and argued in~\cite{Biermann:2010qn} that this is consistent with spectral breaks recently found in hadronic CR data~\cite{Ahn:2010gv,Adriani:2011cu}. Qualitatively, distinguishing between this proposal (two alternative acceleration mechanisms within the same source, which in turn is the high-mass end of the SN sequence) and the above conjecture (``universal'' albeit subtle features of DSA in SNR) might be difficult. One possible feature is that in the Wolf-Rayet scenario there is a potential contribution to positrons from $\beta^{+}$ unstable isotopes produced via photo-dissociation and spallation of the (abundant) nuclei (further considerations on this subject can be found in~\cite{Zirakashvili:2010my}). Hence the normalization of rises in positrons and antiprotons or secondary/primary nuclei might be less sharply linked one another. Of course, a better understanding of additional observables (isotopic anomalies, primary spectral shapes, etc.) may also help clarifying the detailed scenario.
Note that additional sources of $e^{\pm}$ discussed in the literature involve the interaction with dense background {\it photon} densities possibly present near the accelerator~\cite{Stawarz:2009ig,Hu:2009zzb} with more or less reasonable parameters, although it is unclear how the resulting leptons might escape unaffected from the sources while overcoming the strong radiative cooling.

\section{Alternative explanations and contributions from local objects}\label{alternatives}
The well known Shakespeare's quote {\it There are more things in heaven and earth, Horatio, Than are dreamt of in your philosophy} [Hamlet Act 1, scene 5] cannot be more appropriate to
describe the zoo of objects characterizing non-thermal astrophysics. It comes with no surprise that most of them have been associated to leptonic CR spectra. 

For example, some accretion-powered  X-ray binaries made  of a black hole accreting from a stellar companion possess a relativistic jet and act as ``micro-quasars'', small galactic counterparts of their huge cosmological analogues. These objects are known to accelerate leptons at least to GeV energies during flares~\cite{Tavani:2009pm,Hill:2010zx}. According to some optimistic estimates, they may contribute to  a sub-leading yet non-negligible level to the Galactic CR budget~\cite{Heinz:2002qj,Fender:2005nh}. In Ref.~\cite{Ioka:2008cv} such an object  (or alternatively a Gamma-ray burst (GRB) with the right timing/distance) is mentioned as possible source of the anomalies.

Similarly, in~\cite{Kashiyama:2010ui} it was speculated that white dwarf (WD) ``pulsars''\footnote{I.e., a rapidly spinning WD thought to be formed by a merger of two ordinary WDs, since the observed WDs are usually slow rotators.}  could potentially dominate the TeV $e^{\pm}$ window. Although a number of such objects should exist, their ``detection'' as non-thermal sources is still speculative, not to speak of their frequencies and efficiencies as accelerators.
Yet another option investigated in~\cite{Fujita:2009wk} is that positrons originate from collisions of hadronic CRs from a SNR in a {\it nearby} dense cloud environment (see also~\cite{Dog87,Dog90} for similar considerations developed more than two decades ago). In~\cite{Heyl:2010md}, the role of magnetars (ultramagnetized neutron stars) is considered, in addition to ``ordinary'' pulsars.  Needless to say, the list is  incomplete!

From a more general perspective, why such a proliferation of source candidates? There are two reasons for that.
First of all, it has been noted since some time~\cite{Aharonian:1995zz,Atoyan:1996} that at higher and higher energies the contribution of local leptonic CR sources
becomes more and more important, making the approximation of continuous source distribution inappropriate (in particular above ${\cal O}$(100) GeV or so). Turning the
argument around, the TeV range for leptons has a large potential for astrophysical diagnostics, since it may reveal the contribution of single sources, as discussed
e.g. in~\cite{Kobayashi:2003kp}. This can be quickly understood in physical terms: the cooling time $\tau_{\rm cool}=E/(-dE/dt)$ for leptons in the relevant $E$-range is due to radiative losses, and can be estimated in the Thomson limit as
\begin{equation}
\tau_{\rm cool}=\frac{3\,m_e^2\,c^3}{4\,\sigma_T\,u_{\rm tot}E}\approx 10^{15}\,{\rm s}\left(\frac{{\rm eV}/{\rm cm}^3}{u_{\rm tot}}\right)\left(\frac{10\,{\rm GeV}}{E}\right)\,,
\end{equation}
where $u_{\rm tot}$ is the total energy density in low-energy photons (CMB, infrared, optical \ldots) plus the energy density in $B-$field, $u_B=B^2/8\pi$. This implies
an effective ``diffusion horizon'' $R_{\rm diff}$ of the order of 
\begin{eqnarray}
&&R_{\rm diff}\approx 2\sqrt{D(E)\tau_{\rm loss}}\approx\nonumber \\
&& 4\,{\rm kpc }\sqrt{\frac{D_0}{10^{28}\,{\rm cm}^2 {\rm s}^{-1}}\frac{{\rm eV}/{\rm cm}^3}{u_{\rm tot}}}\left(\frac{10\,{\rm GeV}}{E}\right)^\frac{1-\delta}{2}\,,
\end{eqnarray}
i.e. sub-kpc distances for TeV electron sources. Second, the higher in energy the single source starts to contribute, the less demanding the energetic requirement are, although the requirement on $E_{\rm max}$ and the acceleration mechanism tighten.

Given the above arguments,  provided one finds a mechanism to accelerate leptons {\it up to sufficiently high energies and with a sufficiently hard spectrum} (and make them escape!), one can always envisage arranging the distance, time, or normalization of the injection in order to fit the lepton CR data. Of course, when considering known objects whose distance, age and rough energetics is known, as is the case for pulsars, the exercise results significantly more constrained. Many phenomenological attempts have shown that is possible to fit the lepton data up to the highest energies, by adding the contributions of a discrete distribution of pulsars~\footnote{In some cases a lower cut on age is introduced for discarding sources where (presumably) leptons are still confined in the source.}, which can be taken from (admittedly incomplete) pulsar catalogues~\cite{Profumo:2008ms,Grasso:2009ma,Gendelev:2010fd} like the ATNF~\cite{ATNF}, some theoretical models for the distribution of sources in the Galaxy (see e.g.~\cite{Ahlers:2009ae,Delahaye:2010ji}), or a few prominent nearby sources, like Geminga or Monogem in~\cite{Hooper:2008kg,Malyshev:2009tw,Yuksel:2008rf,Grasso:2009ma}.  
Not surprisingly, when fitting the data with one or few pulsars a significantly larger efficiency (typically a few tens of percent) of luminosity conversion into pairs is needed, which is probably also indicative of catalogue incompleteness---at very least for geometric reasons in the case of the pulsar, which can only be discovered when showing the right ``beaming'' towards the Earth.

On the contrary, when invoking putative objects, like a past GRB, the plausibility of the contribution should rather be formulated in terms of probability that the ``right'' distance and timing conditions are realized. For example the proposal formulated in~\cite{Ioka:2008cv} (see also~\cite{Calvez:2010fd}) requires a GRB happening in our Galactic neighborhood in the last $\sim 10^5\div 10^6\,$yr, whose chance probability is probably no higher than $\mathcal{O}(1\%)$ (even forgetting about the evidence that the Milky Way may be too metal-rich to have hosted bright GRBs recently~\cite{Stanek:2006gc}.) An additional aspect is that the cosmic ray spectra produced in (and {\it escaping from}) GRBs, microquasars or more exotic sources are way less constrained, which makes the case for invoking these objects less compelling from the phenomenological point of view, while of course giving more freedom from the astrophysical model-building perspective.

All in all, while  it is not challenging to fit the positron fraction data with a continuous source distribution, see e.g.~\cite{Hooper:2008kg,Kamae:2010ad},
it is very likely that the high-energy part of the lepton CR flux reflects to large extent a few prominent sources; recently,  this has been nicely illustrated in~\cite{Mertsch:2010fn}, with the help of semi-analytical methods and montecarlo simulations (see also~\cite{Kawanaka:2009dk}). A sort of ``galactic variance'' allows one to consider the range $E\gtrsim$TeV for cosmic ray leptons as a window to ``single source contributions": they can certainly be of the same type of those likely to contribute at low energy (like SNRs and PWNe) but possibly---although with smaller probabilities---of a special type.  Clarifying this kind of astrophysical questions is the main purpose of a dedicated experiment, the  ``Calorimetric Electron Telescope'' (CALET)~\cite{Tamura:2010zzc} which is designed to observe electrons up to 20 TeV (and nuclei up to 1000 TeV)
and is expected to be placed at the Japanese Experiment Module at the International Space Station in the current decade.

\section{Perspectives for the future and conclusions}\label{Conclusions}
Compared with the relatively slow progress of the previous decades, the last few years have seen a fast pace of new cosmic ray data. The qualitative improvement
has been particularly spectacular for leptonic CRs. Theoretically, it is challenging if not impossible to explain the data without assuming a primary source of
positrons (or $e^\pm$ pairs). Despite the early enthusiasm about possible dark matter interpretations, it proves very
difficult to reconcile all the cosmic ray data with such an exotic explanation, which (besides having to face theoretical
challenges) is at present even phenomenologically disfavoured. Perhaps more importantly, since at least a couple of decades it has been realized that astrophysical
objects exist which can produce the kind of features presently observed with high statistics, and already hinted to by some past data: even limiting oneself to astrophysical candidates,
numerous options have been suggested.

Here we tried to provide an overview of the situation following a basic criterion: When entering an ``uncharted
territory'' of nature, just like at the beginning of the scientific era, in order to progress in our understanding
it is crucial to make use of Ockham's razor to limit the number of hypotheses, rather than adopting the
strategy to explore all logical possibilities, however tiny their probabilities to be realized in nature are. In fact,
it turns out to be already challenging to explore the consequences of Òplaying with the tools at handÓ, namely to
study more in depth the predicted signals deriving from objects known to exist (and actually being quite common)
in the Galaxy.

This exercise singles out two leading classes of sources, SNRs and PWNe, whose energy budget is
large enough to account for the data and are seen to shine in the non-thermal sky from radio-waves to GeV
and TeV bands. The first problem with SNRs---which are nonetheless the prime candidate as hadronic CR
sources---is that positrons are not present in the nonrelativistic ``background plasma'' into which the shock
front propagates, so they must be secondaries. Inelastic collisions of protons/ions with the background medium may be enough to provide the sufficient number
of positrons, and in some scenarios positrons may also be byproduct of unstable nuclei or reactions onto
photons. A more challenging question is how to get a rising positron fraction rather than a flat ratio at most.
In general, some extra ingredient in the acceleration mechanism is needed, on which different models differ;
also, all models predict correlated ``rising'' signatures in other byproducts like antiproton/proton spectrum or
secondary/primary nuclei. The good news is that forthcoming data from AMS-02 should be able to reveal
these signatures and test the basic idea~\cite{Pato:2010ih}.

Interestingly, even if SNRs turn out not to be  the main contributor to lepton anomalies, it appears reasonable to expect some (in principle measurable) contribution to leptonic and hadronic secondaries at high energy. An important lesson is thus that traditional indicators of DM (like excess of antiprotons) or propagation diagnostics (as both antiprotons and B/C) may have a less trivial energy behaviour than often thought. In a sense, this was already known\footnote{For example, similar considerations had beed done some time ago in the context of cosmic ray acceleration in the interstellar medium~\cite{Cowsik:1980}.}, but somehow forgotten in the last one or two decades when, perhaps in connection with the development of numerical tools for CR propagation (such as GALPROP~\cite{GALPROP}, USINE~\cite{USINE}, DRAGON~\cite{DRAGON}), more attention has been devoted to the uncertainties introduced by the ``propagation'' parameters than those entering the accelerating sources. Fortunately, more and more attention is paid to these issues, as well as to the impact that the primary spectral shape at high energy has on secondaries~\cite{Lavalle:2010sf,Donato:2010vm,Putze:2010fr,Timur:2011vv,Cholis:2011un,Vladimirov:2011rn}.

On the other hand, pulsars (or more exactly PWNe), although traditionally neglected in the CR budget, appear to have both the right energy and the hard spectral shape for leptons at the termination shock to naturally explain the lepton CR anomalies. Yet, the theoretical understanding of acceleration in these objects is less advanced, and some details (as the escape probability in the ISM) appear important to assess quantitatively their contribution to interstellar CR spectra. If they provide the main contribution to the lepton anomalies, it is unlikely that future charged CR data will answer the many obscure points.  More probably, progress will come from improvements into the modeling of their  global emission spectrum from radio to gamma band~\cite{Bucciantini:2010pd,Gelfand:2009aa} as well as from simulations of the microphysics playing a role in these objects, as e.g. in~\cite{Sironi:2009jw,Lyubarsky:2008yi,Sironi:2011zf}.

Whatever the sources of the lepton anomalies are, a dominant contribution of one or few local sources becomes
more and more likely at higher and higher energies. This is a generic prediction of virtually any astrophysical
model which can in principle be confirmed in two ways: i) via high-statistics measurements of the
spectrum, as those expected from CALET~\cite{Tamura:2010zzc}, which
should show ``fine structure'' spectral features due to inhomogeneity
and stochasticity of sources. ii) Via a relatively large anisotropy (roughly at the 0.1\% level at ${\cal O}$(100) GeV), which is  expected from nearby objects~\cite{Buesching:2008hr};  current upper limits from Fermi~\cite{Ackermann:2010ip} appear close to the most optimistic predictions, see~\cite{DiBernardo:2010is,Cernuda:2009kk}.
Some perspectives for AMS-02 in this channel have been studied in~\cite{[113]}.

Let us end this review with a word of caution and a subjective point of view on forthcoming progress.
As usual, one cannot but stress on the importance of additional and high quality data. It is even possible that
current or future experiments might completely revolutionize our understanding of the field, as for example
could follow from the discovery of anti-helium nuclei in AMS-02 or antideuterons (perhaps from DM) in
GAPS~\cite{[114]}. Even lacking such major breakthroughs, it is guaranteed that data from AMS-02 alone will
narrow down uncertainties in the parameters (like primary spectra at high energy, diffusion coefficient, halo
heigth, etc.) entering theoretical predictions of the secondaries~\cite{Pato:2010ih}. However, it is already clear with the current
data that going beyond the zeroth-order description of CR datasets by introducing multiple sources unavoidably
brings huge degeneracies due to the large number of unconstrained parameters, see for example \cite{Delahaye:2010ji}. To
some extent, going beyond the overall distinction between classes of sources (like PWN vs SNR) to the level
of ``inverting'' the problem (i.e. finding actual sources given the spectrum) might make little sense, given the
unavoidable incompleteness of high energy astrophysical catalogues and the fact that, due to diffusive propagation,
some sources contributing now to CR data might not be visible anymore in other bands: the high-energy universe is in fact {\it time-dependent}. 
Thus, it is worth remembering that knowledge in this field does
not automatically follows from the collection of more data: an example is provided by the puzzling observations
of (hadronic) CR multi-TeV anisotropy from many experiments \cite{[115],[116],[117],[118],[119],[120]}, for which
no convincing explanation has been found, yet. To put that in the perspective of our original problem of lepton
diagnostics, detecting anisotropy might not be enough to make source identification, since other effects (like inhomogeneous diffusion coefficient, anisotropic diffusion,
local ÒbubblesÓ, etc.) may even mask or alter the source effect. Perhaps energy-dependent anisotropy studies
might be needed for diagnostics, but these appear even
more-challenging.
If compared with the efforts gone into modelling cosmic
ray propagation or astrophysical models for photon astronomy, the theoretical understanding of high energy
sources of the cosmic rays eventually detected
at the Earth (i.e. not remotely!) is much more rudimentary,
also because of the highly indirect nature of the
problem. It is auspicable that the excitement brought in
the astroparticle community by the recent data will also
stimulate a new generation of scientists to devote themselves
to theoretical aspects of acceleration and escape
of CRs from astrophysical sources, since firm progress
in solving these fascinating questions will certainly require
additional ideas and tools and a critical re-examination
of the many simplifying assumptions underlying
present scenarios.

\section*{Acknowledgments}
I would like to thank F. Donato, N. Fornengo,  and P. Salati for comments on the manuscript and all my past collaborators,
in particular P. Blasi, for endless and fruitful discussions on the topics covered by this article.

\bibliographystyle{elsarticle-harv}

\end{document}